\documentclass[11pt]{article}
\usepackage{amssymb}

\newcommand{\half}{\mbox{$\frac{1}{2}$}}
\newcommand{\quart}{\mbox{$\frac{1}{4}$}}

\newcommand{\Tr}{\mbox{Tr}}

\begin{document}

%\begin{table}
%\begin{flushright}
%UIO
%\end{flushright}
%\end{table}

\title{Radiative corrections to the Casimir energy \\
and effective field theory}

\author{Finn Ravndal\thanks{E-mail: {\tt finn.ravndal@fys.uio.no}} \ and
Jan B.\ Thomassen\thanks{E-mail: {\tt thomasse@kph.tuwien.ac.at}} \\
{\em Institute of Physics, University of Oslo} \\ {\em N--0316 Oslo,
Norway}}

\date{February 21, 2001}

\maketitle

\begin{abstract}

We discuss radiative corrections to the Casimir effect from an
effective field theory point of view. It is an improvement and more
complete version of a previous discussion by Kong and Ravndal. By
writing down the most general effective Lagrangian respecting the
symmetries and the boundary conditions, we are able to reproduce
earlier results of Bordag, Robaschik and Wieczorek calculated in full
QED. They obtained the correction $E^{(1)}_0 = \pi^2\alpha/2560mL^4$
to the Casimir energy. We find that this leading correction is due to
surface terms in the effective theory, which we attribute to having
dominant fluctuations localized on the plates.

\vspace{\baselineskip}

\end{abstract}

\section{Introduction}

Although the Casimir effect \cite{casimir,itzykson} has been known for
more than 50 years, the question of what are the leading quantum
corrections to this effect is surprisingly still a subject of
debate. The corrections we have in mind here are those that are caused
by the coupling of the electromagnetic field to the electron
field. These are first of all important for theoretical reasons, but
recent improvements in experimental techniques \cite{lamoreaux} may
lead to interesting confrontations between theory and experiment in
the near future.

The first attempts to determine the quantum corrections to the Casimir
energy $E_0$ were reported in a paper by Bordag, Robaschik, and
Wieczorek \cite{bordag} (BRW). These authors considered the quantum
vacuum within the usual set-up with two perfectly conducting parallel
plates using full QED. The electromagnetic field satisfies metallic
boundary conditions, while the electron field does {\em not}
feel the presence of the metallic plates. They found the correction
$E^{(1)}_0 = \pi^2\alpha/2560mL^4$ to the well-known leading term
$E^{(0)}_0 = -\pi^2/720L^3$, where $L$ is the separation between the
plates and $m$ is the electron mass. This correction emerges as an
effect of vacuum polarization. Three years ago, one of us and Kong
\cite{kong} (KR) studied the Casimir effect from an effective field
theory \cite{eft} point of view. In that work it was argued that
vacuum polarization does not have any effect on the Casimir
energy. This would be just like for effective QED in free space where
it is known that vacuum polarization has no physical consequence in
the absence of external electrons. The leading corrections would then
come from the Euler--Heisenberg effective Lagrangian and have the
value $E^{(1)}_0 = 11\pi^4\alpha^2/2^7 3^5 5^3m^4L^7$, in disagreement
with the result of BRW. Subsequently, the use of effective QED in KR
was criticized in Ref.\ \cite{bordag-sch}. The authors of
\cite{bordag-sch} pointed out that even though the Casimir effect is a
low energy phenomenon, a derivative expansion, as is typical of
effective theories, could not be used in this case. The reason for
this is that the evaluation of the corrections to the Casimir energy
in the full theory involves an integral along the cut of the vacuum
polarization tensor $\Pi_{\mu\nu}(k^2)$, and this information is lost
if $\Pi_{\mu\nu}$ is expanded in powers of $k^2$. This might then
``explain'' why the results in KR were incorrect and indicate that the
result of BRW was correct in the first place.

Thus, Ref.\ \cite{bordag-sch} gave the impression that effective field
theory methods could not be used on the Casimir problem. However,
effective field theory is not just a derivative expansion of the
underlying theory but something more general. Indeed, effective field
theory {\em always} works when the physical degrees of freedom are
fields. This is because the effective theory is {\em by definition}
constructed to give the same results as the underlying theory in a
given situation, usually at low energies. There is therefore no reason
to dismiss the results from KR on the grounds that they are based on
effective field theory methods. We are in fact left with the following
possibilities: (a) The full QED result of BRW is correct and the
effective QED discussion in KR is incomplete and for this reason gives
the wrong answer. (b) The full QED calculation in BRW is incorrect and
the results from effective QED in KR are correct. (c) Both the full
and the effective QED results are wrong.

In this paper, we show that it is possibility (a) above that is
correct by explicitly constructing the effective theory. We start in
Sec.\ 2 by recalling how perturbation theory works in the presence of
boundary conditions, as first discussed in BRW. We also sketch the
calculation of the radiative corrections to the Casimir energy.

In Sec.\ 3 we turn to the construction of the effective theory. All
possible terms of the Lagrangian that respect the symmetries must be
written down. In this process we realize that there is a class of
terms that were left out in the previous treatment of effective QED in
KR. These are surface terms, where the contributions to the action
live only on the two plates. We then establish the counting rules of
the effective theory, and it is seen that the surface terms provide
the leading correction to the Casimir energy.

In Sec.\ 4 we use this effective theory to calculate the Casimir
energy corrections in terms of the low energy constant of the leading
surface term in the effective Lagrangian. We are then able to
determine this constant by matching to the BRW result of the full
theory. This calculation demonstrates that effective field theory can
be applied to the Casimir effect. However, more constants must be
determined by matching before we can claim to have a workable
effective theory with predictive power.

Sec.\ 5 contains a brief discussion of our results. In particular,
we demonstrate one possible way of understanding the relevant surface
terms as being ``generated'' from the full theory in the low energy
approximation.

\section{Functional derivation of the Casimir energy}

In this section we make some general remarks and recall how the photon
propagator is modified in the presence of the two plates, which allows
perturbation theory to work the usual way. We also briefly sketch the
calculation of the radiative corrections to the Casimir energy. There
are no new results in this section.

The starting point is the path integral over the photon field $A_\mu$,
\begin{eqnarray}
Z & = & \int{\cal D}Ae^{iS},
\end{eqnarray}
with the Maxwell action $S=\int d^4x(-\frac{1}{4}F_{\mu\nu}
F^{\mu\nu})$. It is to be understood that only configurations
respecting the boundary conditions are integrated over. The system we
consider will always be the standard geometry of two perfectly
conducting parallel plates positioned at $z=0$ and $z=L$. Hence the
electromagnetic field respects the usual metallic boundary conditions
$\mathbf{n\times E = n\cdot B} = 0$ at the plates, where
$\mathbf{n}=(0,0,1)$ is the normal vector to the plates in the
positive $z$-direction. In four-vector notation the boundary condition
reads $n^\mu\tilde F_{\mu\nu}|_{z=0,L}=0$, where $\tilde F_{\mu\nu} =
\half\epsilon_{\mu\nu\rho\sigma}F^{\rho\sigma}$ is the dual field
strength tensor.

In order to make the boundary conditions explicit in the path
integral, we may introduce a product of two delta functions enforcing
these boundary conditions. At the same time we extend the integrations
to {\em all} configurations of $A_\mu$:
\begin{eqnarray}
Z & = & \int{\cal D}A\delta(n^\mu\tilde F_{\mu\nu}|_{z=0})
\delta(n^\mu\tilde F_{\mu\nu}|_{z=L})e^{iS}.
\end{eqnarray}
We then represent the delta functions by path integrals over two
external fields $B^i_\mu(x_\perp)$, $i=1,2$, $x^\mu_\perp =
(x^0,x^1,x^2)$, that lives only on the two plates at $z=a_i$ with
$a_1=0$, $a_2=L$. The complete path integral thus becomes
\begin{eqnarray}
\label{modified-z}
Z & = & \int{\cal D}A{\cal D}Be^{iS'},
\end{eqnarray}
with the modified action
\begin{eqnarray}
\label{modified-action}
S' & = & -\quart\int d^4xF_{\mu\nu}(x)F^{\mu\nu}(x)
- \int d^3x_\perp B_i^\mu(x_\perp)n^\mu\tilde F_{\mu\nu}(x_\perp,a_i).
\end{eqnarray}
A summation over $i$ is understood in this expression.

Apart from these constraints, the fields are assumed to exist both
outside as well as in between the plates. This set-up is of course
highly idealized, and in order to make the description more realistic
we need, at least, to take into account the physical nature of the
plates and the coupling of $A_\mu$ to the electron field. In this
paper we will consider corrections from the electron field, but where
this electron field does not feel any boundary conditions at the
plates. This is the same situation as in BRW and \cite{bordag-lin},
and can be considered as a first step on the way to a more realistic
treatment of corrections to the Casimir energy.

In BRW it was pointed out that standard Feynman perturbation theory
can be used even in the presence of the metallic boundary conditions
provided the correct modified photon propagator is used. An expression
for this propagator may be obtained by first coupling external sources
to the photon field in the action (\ref{modified-z}). Integrating out
the photon field and the Lagrange multiplier fields $B_\mu^i$ thus
produces a functional of the external sources from which the
propagator can be read off. The result is
\begin{eqnarray}
\label{propagator}
\langle 0|TA_\mu(x)A_\nu(x')|0\rangle & = & iD_{\mu\nu}(x-x')
-i\bar D_{\mu\nu}(x,x'),
\end{eqnarray}
where $D_{\mu\nu}(x-x')$ is the free propagator and $\bar
D_{\mu\nu}(x,x')$ is
\begin{eqnarray}
\nonumber
\bar D_{\mu\nu}(x,x') & \equiv & \int\frac{d^3k_\perp}{(2\pi)^3}
\frac{-P_{\mu\nu}^\perp}{4\gamma\sin\gamma L}
e^{-ik_\perp(x_\perp-x'_\perp)} \\
\nonumber
  & & \mbox{} \hspace{2em} \times\left[e^{-i\gamma L}
\left(e^{i\gamma|z|}e^{i\gamma|z'|}
+e^{i\gamma|z-L|}e^{i\gamma|z'-L|}\right)\right. \\
  & & \mbox{} \hspace{5em} \left.-\left(e^{i\gamma|z|}
e^{i\gamma|z'-L|}
+e^{i\gamma|z-L|}e^{i\gamma|z'|}\right)\right].
\end{eqnarray}
The notation here is $k_\perp^\mu = (k^0,k^1,k^2)$, $\gamma =
\sqrt{k^2_\perp} = \sqrt{k_0^2-k_1^2-k_2^2}$, and $P_{\mu\nu}^\perp$
is the projection operator
\begin{eqnarray}
P_{\mu\nu}^\perp & = & \left\{ \begin{array}{ll}
g_{\mu\nu}-\frac{k_\mu k_\nu}{k_\perp^2}
  & \mbox{for $\mu,\nu\neq 3$,} \\
0 & \mbox{for $\mu=3$ or $\nu=3$.} \end{array} \right.
\end{eqnarray}
This modified propagator can then be used for calculating diagrams in
perturbation theory when the interactions with the electrons are
turned on. The correction to the Casimir energy can be found in this
way by computing the relevant diagrams.

For later convenience we now briefly sketch how the corrections may be
calculated. Our approach is essentially that of \cite{bordag-lin},
which is an improved calculation compared to the original one in
BRW. We are also inspired by the discussion in \cite{kardar} which is
another approach based on the functional formalism. The general idea
is to make use of the field theory identity
\begin{eqnarray}
\label{identity}
Z & = & e^{-iE_\mathrm{vac}T},
\end{eqnarray}
where $E_\mathrm{vac}$ is the vacuum energy and $T$ is the total
time. The Casimir energy $E_0$ is defined to be $E_0 =
E_\mathrm{vac}/A$ where $A$ is the area of the plates. We may thus
extract $E_0$ by evaluating a path integral.

We will first consider the simpler case of the plain Casimir effect
without radiative corrections before we consider the full complexity
of the problem. If we define a new field $C_\mu^i \equiv
-n^\alpha\epsilon_{\alpha\mu\nu\beta}\partial^\beta B_i^\nu$, we can
write the second term in the action (\ref{modified-action}) as $-\int
d^3x_\perp A_\mu P^{\mu\nu}_\perp C_\nu^i$, where we also use the
projection operator $P^\perp_{\mu\nu}$. Now, since the photon field
$A_\mu$ couples directly to this field, we can without loss of
generality change path integration variables from $B_i^\mu$ to
$C_i^\mu$ as long as we make sure that $C_i^\mu$ has the properties
$\partial_\mu C^\mu_i=0$ and $C^3_i=0$. The new path integral over
$C_i^\mu$ still imposes the same boundary conditions on $A_\mu$. The
two conditions on $C_i^\mu$ means that this field has only two
independent components, which implies that only two components of
$A_\mu$ couples to $C_i^\mu$. The other two components thus decouples
from the problem and has no physical consequence. On the other hand
$A_\mu$, or more generally any vector field, can be decomposed into
$A_\mu = P_{\mu\nu}^\perp A_\nu + A'_\mu$, where the two terms are
orthogonal to each other. We can use this to rewrite the kinetic term
of the photons: $-\frac{1}{4}F_{\mu\nu}F^{\mu\nu} = \half A_\mu\square
P^{\mu\nu}_\perp A_\nu + \mbox{(terms with $A'_\mu$)}$, where we have
used partial integration and the fact that
$P_{\mu\nu}^\perp(g^{\nu\rho} - \partial^\nu\partial^\rho/\square)
P_{\rho\sigma}^\perp = P_{\mu\sigma}^\perp$. The result of all this is
that we are allowed to write the path integral over $A_\mu$ and
$C^i_\mu$ with the action
\begin{eqnarray}
S'' & = & \half\int d^4xA_\mu(x)\square P^{\mu\nu}_\perp A_\nu(x)
-\int d^3x_\perp A_\mu(x_\perp,a_i)P^{\mu\nu}_\perp C^i_\nu(x_\perp).
\end{eqnarray}
The terms depending on $A'_\mu$ have been omitted since they decouple
from $A_\mu$ and $C_\mu^i$ and are therefore irrelevant to the Casimir
energy. Note also that it has not been necessary to consider any
specific gauge in order to arrive at this.

This path integral is Gaussian in $A_\mu$ and may be
evaluated to give a path integral over $C_\mu^i$ alone with the new
action
\begin{eqnarray}
S & = & -\half\int d^3x_\perp C_\mu^i
M_{ij}P^{\mu\nu}_\perp C^j_\nu,
\end{eqnarray}
where $M_{ij}$ is the operator whose form in momentum space is
\begin{eqnarray}
M_{ij} & = & \frac{i}{2\gamma}e^{i\gamma|a_i-a_j|}
  \;=\; \frac{i}{2\gamma}
\left(\begin{array}{cc} 1 & e^{i\gamma L} \\
e^{i\gamma L} & 1 \end{array}\right).
\end{eqnarray}

The path integral then becomes
\begin{eqnarray}
Z & = & (\mathrm{Det}\;P^{\mu\nu}_\perp M_{ij})^{-1/2}
  \;=\; (\mathrm{Det}\;M)^{-1}
  \;=\; e^{-\mathrm{Tr}\ln\det M},
\end{eqnarray}
where '$\mathrm{Det}\;M$' means determinant with respect to both
$k$-space and $\mathit{ij}$-indices, while '$\det M$' means
determinant with respect to the $\mathit{ij}$-structure only. We have
also used that $P^{\mu\nu}_\perp$ is a projection operator onto a
two-dimensional space and thereby gives a multiplicity $2$. Therefore,
\begin{eqnarray}
\label{energy-c}
E_0 & = & -\frac{i}{AT}\Tr\ln\det M.
\end{eqnarray}
It is straightforward to evaluate this expression in momentum space,
where $\Tr \to AT\int d^3k_\perp/(2\pi)^3$, and
\begin{eqnarray}
\det M & = & \frac{-1}{4\gamma^2}\left(1-e^{2i\gamma L}\right).
\end{eqnarray}
When we take the logarithm of this, the term with $\ln(-1/4\gamma^2)$
does not depend on $L$ and can be ignored in this context. Thus we
find that the Casimir energy is
\begin{eqnarray}
\nonumber
E_0 & = & -i\int\frac{d^3k_\perp}{(2\pi)^3}
\ln\left(1-e^{2i\gamma L}\right)
  \;=\; -\frac{\pi^2}{720L^3},
\end{eqnarray}
as it should be.

Let us now turn to the quantum corrections to this result. As is
well-known, the full effect that the electrons have on the vacuum
polarization to order $\alpha$ can be summarized by replacing the
kinetic term of free photons with a modified Lagrangian: $-\quart
F_{\mu\nu}F^{\mu\nu} \to -\quart F_{\mu\nu}[1 +
\Pi(-\square)]F^{\mu\nu}$, where $\Pi(-\square)$ is the usual
renormalized vacuum polarization modulo the gauge invariant projection
operator $\square g_{\mu\nu} - \partial_\mu\partial_\nu$.

This leads to a modification of the photon propagator and in turn this
leads to a modification of $M_{ij}$ and thereby its determinant, which
now becomes
\begin{eqnarray}
\det M & = & \frac{-1}{4\gamma^2}
\left(1-e^{2i\gamma L}\right)
\left(1-4i\int_{-\infty}^\infty\frac{dk_z}{2\pi}\frac{\Pi(k^2)}{k^2}
\frac{\gamma(1-e^{i\gamma L}\cos k_zL)}{1
-e^{2i\gamma L}}\right).
\end{eqnarray}
When the logarithm of this expression is inserted in formula
(\ref{energy-c}) for the Casimir energy we recognize the first
parenthesis as the contribution giving the leading term $E_0^{(0)} =
-\pi^2/720L^3$ calculated above. The second parenthesis then gives the
correction:
\begin{eqnarray}
\label{brw-result}
E_0^{(1)} & = & -2i\int\frac{d^4k}{(2\pi)^4}
\frac{\gamma}{\sin\gamma L}
\left(e^{-i\gamma L}-\cos k_zL\right)\frac{\Pi(k^2)}{k^2},
\end{eqnarray}
which is identical to Eq.\ (45) in \cite{bordag-lin}, apart
from a difference in sign due to different sign conventions for
$\Pi(k^2)$. It was evaluated there in the physically interesting limit
where $mL\gg 1$:
\begin{eqnarray}
E_0^{(1)} & = & \frac{\pi^2\alpha}{2560mL^4}.
\end{eqnarray}
This is the result that was first obtained by BRW.

\section{Effective QED with two conducting plates}

We now turn to effective field theory \cite{eft}. An effective field
theory calculation starts with writing down the most general effective
Lagrangian respecting the symmetries of the problem. Next, one assigns
counting rules to each term in the effective Lagrangian, which allows
us to calculate physical quantities from the effective theory in a
systematic way. The free coefficients that multiplies each term in the
effective Lagrangian -- the ``low energy constants'' -- are then
determined by matching the results with the corresponding quantities
calculated in the full theory. The result is a theory that in
principle may be used to perform calculations to any desired order in
the effective counting rules. This program is applied here to the
Casimir energy which enables us to explicitly demonstrate that
effective field theory works in this case.

Let us begin our investigations by writing down the effective
Lagrangian, and then discuss it afterwards. It has the form
\begin{eqnarray}
\label{l-eff}
\nonumber
{\cal L}_\mathrm{eff} & = & {\cal L}^\mathrm{bulk}_\mathrm{eff}
+ {\cal L}^\mathrm{surf}_\mathrm{eff}, \\
\nonumber
{\cal L}^\mathrm{bulk}_\mathrm{eff}
  & = & -\quart F_{\mu\nu}F^{\mu\nu}
+ \frac{c_1}{m^2}F_{\mu\nu}
n^\rho\partial_\rho n^\sigma\partial_\sigma F^{\mu\nu}
+ \frac{c_2}{m^2}F_{\mu\nu}n^\mu\partial^\nu
n^\rho\partial^\sigma F_{\rho\sigma} \\
\nonumber
  & & \mbox{} +\frac{c_3}{m^2}F_{\mu\nu}\square F^{\mu\nu}
+ \mbox{higher orders}, \\
\nonumber
{\cal L}^\mathrm{surf}_\mathrm{eff}
  & = & -\frac{d_1}{4m}F_{\mu\nu}F^{\mu\nu}
[\delta(z)+\delta(z-L)] + \mbox{higher orders}.
\end{eqnarray}
The terms fall into two categories, a ``bulk'' contribution where
field values from the whole of space-time are used and a ``surface''
contribution where only field values at the two plates at $z=0$ and
$z=L$ are used. It is to be understood that the metallic boundary
conditions at the plates are accounted for by inserting delta
functions in the path integral where $\int d^4x{\cal L}_\mathrm{eff}$
is the action.

But let us first comment on the symmetries of the problem. In free
space, these would be gauge symmetry, Lorentz symmetry, translation
symmetry and the discrete symmetries. In the presence of the two
plates, however, parts of the Lorentz and translation symmetries are
broken. Hence more terms are allowed in the effective Lagrangian. We
can include some of these terms in a mock Lorentz invariant form by
making use of the four-vector $n^\mu$ from the last Section. Due to
gauge invariance we are required to build the effective theory from
$F_{\mu\nu}$ rather than from $A_\mu$. Thus, using $F_{\mu\nu}$'s,
$n$'s and derivatives we get an effective Lagrangian of the form
(\ref{l-eff}). Note that there can be no terms with an odd number of
$n$'s. The reason for this is that $n^\mu$ also appears in the delta
function that enforce the proper boundary conditions in the {\em full}
theory in the previous Section. It is then evident that there is an
additional symmetry, $n^\mu \to -n^\mu$, which rules out odd numbers
of $n$'s.

The surface terms in ${\cal L}_\mathrm{eff}^\mathrm{surf}$ deserves
special mention. They must {\em a priori} be present in the action
because they respect the symmetries. Let us also remember that surface
terms are not unknown in the world of effective theories. The
Wess--Zumino--Witten term in chiral perturbation theory has such an
interpretation \cite{witten}. There is also a relation to the
hyperfine splitting term, proportional to a delta function, in the
effective Hamiltonian that is often used in the discussion of
hydrogen-like atoms \cite{itzykson}.

The effective Lagrangian in this problem is given in a natural way in
terms of two derivative expansions -- one in the bulk and one on the
plates. The coefficient in front of each term has been scaled with the
electron mass $m$ so that the parameters $c_i$ and $d_i$ are
dimensionless. The other scale in the theory, $L$, is not expected to
have any influence on a {\em local} quantity like the effective
Lagrangian. Only terms up to mass dimension six for the bulk part, and
up to dimension five for the surface part, are displayed in Eq.\
(\ref{l-eff}). It may appear that a third term with two $n$'s could be
written down in ${\cal L}_\mathrm{eff}^\mathrm{bulk}$, proportional to
$F_{\mu\nu} n^\mu\partial^\rho n^\sigma\partial_\sigma {F_\rho}^\nu$.
However, since the boundary conditions are taken care of by two delta
functions inserted in the path integral, we are allowed to freely
perform partial integrations. One may then check that this interaction
is equivalent to the term with $c_1$ in (\ref{l-eff}). It may also
appear that another term proportional to $n_\mu F^{\mu\rho}n^\nu
F_{\nu\rho}[\delta(z)+\delta(z-L)]$ could be written down in ${\cal
L}^\mathrm{surf}_\mathrm{eff}$, but due to the boundary conditions
this term is equivalent to the one already given.

In order to assign counting rules for the effective theory we start
with the ``free'' term $-\quart F_{\mu\nu}^2$. We may arbitrarily
assign the order $p^2$ to this, where $p$ is a ``small''
momentum. Then $F_{\mu\nu}$ is of order $p$. Likewise we assign a
factor $p$ for each derivative, so that the displayed terms with $c_i$
in ${\cal L}_\mathrm{eff}^\mathrm{bulk}$ are of order
$p^4$. Furthermore, it is natural to assign a factor of $p$ to the
$\delta$-functions in ${\cal L}_\mathrm{eff}^\mathrm{surf}$ in
agreement with dimensional analysis. Thus, the leading surface term,
proportional to $d_1$, is of order $p^3$. This means that unless $d_1$
accidentally vanishes or is unnaturally small, the leading corrections
to the Casimir energy in the effective theory comes from this term.

The order $p^4$ operators we have written in the effective Lagrangian
(\ref{l-eff}) are all terms that vanish in free space. Indeed, it is
well-known in free space effective field theory that terms in the
action which vanishes due to the free field equations of motion can be
removed by a field redefinition in the path integral
\cite{georgi-npb}. This would remove the term proportional to $c_2$ in
Eq.\ (\ref{l-eff}). On the other hand, the Uehling term proportional
to $c_3$ may be removed by the transformation
\begin{eqnarray}
\label{A-shift}
A_\mu & \to & A_\mu - \frac{2c_3}{m^2}\square A_\mu.
\end{eqnarray}
However, this possibility is no longer open to us in the present case
due to the nontrivial boundary conditions. The point is that even
though one field configuration satisfies the required condition on the
plates, this will in general not be true for the transformed
configuration because of the presence of derivatives in the
transformation law. This is not a problem in free space, since it is
always implicitly understood that both the fields and their
derivatives go to zero at infinity. Therefore, the Uehling term and
other terms in the same situation must be kept and may {\em a priori}
give rise to real physical effects.

The contribution to the Casimir energy from the various correction
terms in (\ref{l-eff}) can be obtained from the field theory identity
(\ref{identity}). From this we find the correction
\begin{eqnarray}
\label{e-casimir}
E^{(1)}_0 & = & -\frac{1}{AT}\langle S^{(1)}\rangle,
\end{eqnarray}
where $S^{(1)}$ is a small perturbation of the leading order Maxwell
action $S^{(0)}$. The expectation values in $\langle\cdots\rangle$
refers to the theory described by $S^{(0)}$ under the further
restriction that the usual metallic boundary conditions $n^\mu\tilde
F_{\mu\nu} = 0$ hold on the plates. This means that the contractions
that appear in $\langle\cdots\rangle$ should be calculated using the
modified photon propagator (\ref{propagator}) of BRW, as discussed in
Sec.\ 2.

\section{Corrections to the Casimir energy in the effective theory}

The leading order correction is expected to result from the lowest
dimension operator in the Lagrangian (\ref{l-eff}), {\em i.e.}
\begin{eqnarray}
\Delta{\cal L} & = & -\frac{d_1}{4m}F_{\mu\nu}F^{\mu\nu}
[\delta(z)+\delta(z-L)].
\end{eqnarray}
In order to make the contractions involved in calculating
$\langle\Delta S\rangle$, it is appropriate to write $\Delta S$ so
that it depends directly on $A_\mu$ instead of $F_{\mu\nu}$. For the
plate at $z=a$, with $a$ either $0$ or $L$, we have
\begin{eqnarray}
\Delta S_a & = & -\frac{d_1}{4m}\int d^4x
F_{\mu\nu}F^{\mu\nu}\delta(z-a).
\end{eqnarray}
We may perform partial integration which gives
\begin{eqnarray}
\label{4d-form}
\Delta S_a & = & \frac{d_1}{2m}\int d^4xA_\mu\left[\delta(z-a)\square
+\half\delta''(z-a)\right]A^\mu,
\end{eqnarray}
where $\delta''(z-a) = (d^2/dz^2)\delta(z-a)$. In this expression we
have omitted terms that do not contribute to the result. Indeed, when
(\ref{4d-form}) is contracted with the modified photon propagator,
only the part with $\bar D_{\mu\nu}(x,x')$ carries
$L$-dependence. This term involves the projection operator
$P_{\mu\nu}^\perp$ so that, without loss of generality, we may make
the replacement
\begin{eqnarray}
\half A_\mu[\cdots]A^\mu & \to & \half P^{\mu\nu}_\perp
A_\mu[\cdots]A_\nu.
\end{eqnarray}
The correction then becomes
\begin{eqnarray}
\label{casimir-corr}
E^{(1)}_0 & = & -\frac{1}{AT}\frac{d_1}{2m}\int d^4x\sum_{i=1}^2
P^{\mu\nu}_\perp\langle A_\mu(x)[\delta(z-a_i)\square
+\half\delta''(z-a_i)]A_\nu(x)\rangle.
\end{eqnarray}

Let us first consider the term involving $\delta''$ and set
$i=1$. From Eq.\ (\ref{propagator}) it then follows
\begin{eqnarray}
\label{delta''}
\langle A_\mu(x)\delta''(z)A_\nu(x)\rangle
  & = & \delta''(z)i(D_{\mu\nu}(x,x)-\bar D_{\mu\nu}(x,x)).
\end{eqnarray}
Since only $\bar D_{\mu\nu}$ contains dependence on $L$ we can
disregard the part involving $D_{\mu\nu}$. The right-hand side of
(\ref{delta''}) then becomes
\begin{eqnarray}
-i\int\frac{d^3k_\perp}{(2\pi)^3}
\frac{-P_{\mu\nu}^\perp\delta''(z)}{4\gamma\sin\gamma L}
\left[e^{-i\gamma L}(e^{2i\gamma|z|}+e^{2i\gamma|z-L|})
-2e^{i\gamma(|z|+|z-L|)}\right],
\end{eqnarray}
which appears under a space-time integral. We may therefore partially
integrate to remove the two $z$-derivatives on the
$\delta$-function. Making use of identities such as
\begin{eqnarray}
\nonumber
\partial_ze^{i\gamma|z-a|}
  & = & i\gamma\epsilon(z-a)e^{i\gamma|z-a|}, \\
\partial_z^2e^{i\gamma|z-a|}
  & = & 2i\gamma\delta(z-a)-\gamma^2e^{i\gamma|z-a|},
\end{eqnarray}
where $\epsilon(x)$ is the sign function, this leads to
\begin{eqnarray}
\langle A_\mu(x)\delta''(z)A_\nu(x)\rangle
  & = & -iP_{\mu\nu}^\perp\delta(z)\int\frac{d^3k_\perp}{(2\pi)^3}
\frac{\gamma e^{i\gamma L}}{\sin\gamma L}.
\end{eqnarray}
Similarly, the term with $i=2$ gives
\begin{eqnarray}
\langle A_\mu(x)\delta''(z-L)A_\nu(x)\rangle
  & = & -iP_{\mu\nu}^\perp\delta(z-L)\int\frac{d^3k_\perp}{(2\pi)^3}
\frac{\gamma e^{i\gamma L}}{\sin\gamma L}.
\end{eqnarray}
On the other hand, the terms in (\ref{casimir-corr}) involving
$\delta\square$ are found to be equal to
\begin{eqnarray}
\langle A_\mu(x)\delta(z-a_i)\square A_\nu(x)\rangle
  & = & iP_{\mu\nu}^\perp\delta^4(0)\delta(z-a_i),
\end{eqnarray}
which do not give any $L$-dependent contribution to the correction
$E^{(1)}_0$.

Collecting this information we find that the contribution to the
Casimir energy is
\begin{eqnarray}
\label{delta-e}
\nonumber
E^{(1)}_0 & = & -2\frac{d_1}{2m}P^{\mu\nu}_\perp
\half(-iP_{\mu\nu}^\perp)\int\frac{d^3k_\perp}{(2\pi)^3}
\frac{\gamma e^{i\gamma L}}{\sin\gamma L} \\
\nonumber
  & = & \frac{d_1}{m}i\int\frac{d^3k_\perp}{(2\pi)^3}
\frac{\gamma e^{i\gamma L}}{\sin\gamma L} \\
\nonumber
  & = & -\frac{d_1}{2\pi^2mL^4}
\int_0^\infty dk\frac{k^3e^{-k}}{\sinh k} \\
  & = & -\frac{\pi^2d_1}{240mL^4},
\end{eqnarray}
where we have used that $P_{\mu\nu}^\perp P^{\mu\nu}_\perp=2$. Thus,
by choosing the numerical value
\begin{eqnarray}
d_1 & = & -\frac{3\alpha}{32}
\end{eqnarray}
we are able to reproduce the full QED result $E^{(1)}_0 =
\pi^2\alpha/2560mL^4$. This is a reasonable magnitude for a constant
that represents radiative corrections from virtual electron-positron
pairs.

\section{Discussion}

To summarize, we have constructed an effective field theory
satisfying the appropriate boundary conditions, and have been able to
reproduce the full QED result to next-to-leading order of the low
energy expansion. This demonstrates that effective field theory works
for the Casimir effect -- as it does for any other field theoretical
problem.

The Lagrangian of an effective field theory is constructed by writing
down all possible terms that respect the symmetries and the
coefficients that multiplies them are determined by a matching
procedure. The question therefore never arises in an effective field
theory calculation how the various terms in the effective Lagrangian
``comes about''. Nevertheless, it may be interesting in the present
case to see if we can gain some insight into the nature of the surface
terms that are so important for our results. Referring back to Eq.\
(\ref{brw-result}) we have the following expression for the
energy correction \cite{bordag},
\begin{eqnarray}
\label{correction}
E_0^{(1)} & = & -2i\int\frac{d^4k}{(2\pi)^4}
\frac{\gamma}{\sin\gamma L}
\left(e^{-i\gamma L}-\cos k_zL\right)\frac{\Pi(k^2)}{k^2},
\end{eqnarray}
in the underlying theory. If now we are interested in the low energy
content of this expression, we may try to simplify it by taking the
low energy limit of $\Pi$:
\begin{eqnarray}
\Pi(k^2) & = & c\frac{k^2}{m^2} + {\cal O}(k^4),
  \hspace{2em} \mbox{for $k^2\ll m^2$},
\end{eqnarray}
with $c$ some unimportant constant. Inserting this in
Eq.\ (\ref{correction}), we find that the term proportional to $\cos
k_zL$ vanishes from the $k_z$-integration, and we get
\begin{eqnarray}
E^{(1)}_0 & = & -2i\frac{c}{m^2}\int\frac{d^4k}{(2\pi)^2}
\frac{\gamma e^{i\gamma L}}{\sin\gamma L} + {\cal O}(k^4).
\end{eqnarray}
Here we have used that $e^{-i\gamma L} = e^{i\gamma L} - 2i\sin\gamma
L$ and that the $\sin\gamma L$-part of this leads to an
$L$-independent contribution. The $k_z$-integration here is
divergent. However, we should remember that this form is valid only
for $k^2\ll m^2$. We can repair the situation by choosing a cutoff
for the $k_z$-integration of the order $m$. We then get
\begin{eqnarray}
E^{(1)}_0 & \sim & \frac{i}{m}\int\frac{d^3k_\perp}{(2\pi)^3}
\frac{\gamma e^{i\gamma L}}{\sin\gamma L} + \cdots
\end{eqnarray}
for the energy correction. On comparison, we see that this has the
same form as the expression in the second line of Eq.\
(\ref{delta-e}), which is precisely the contribution from the surface
term.

It is also possible to understand how the cut of the vacuum
polarization tensor, with modes of relatively high momentum, $k^2\geq
4m^2$, may contribute to the full QED energy correction. At first
sight it would appear that the effects of virtual electron-positron
pairs with high energies are unimportant to the Casimir problem. After
all, such pairs are sharply localized while the Casimir effect is
controlled by the macroscopically large scale $L$, the distance
between the plates. However, it is known from the literature on the
Casimir effect that the fluctuations of the electromagnetic field are
large near the plates and diverge as we approach them. In this region
with violent fluctuations there will be an increase in the production
of pairs. Thus, even sharply localized pairs may be expected to
contribute significantly to the correction. This picture of what is
going on means that part of the physics is localized on the plates and
agrees very well with the presence of surface terms in the effective
theory.

As mentioned in Sec.\ 3, the calculations we have done in this paper
are only the first steps in a full-fledged effective field theory
calculation. At higher orders of the counting rules more terms with
their corresponding low energy constants enters the description and
these constants must also be determined by matching. It will then be
necessary to consider other quantities than the Casimir energy like,
for example, the Green's functions of the system. This would then
allow us to perform more detailed calculations in the effective theory
to any desired order of precision in the low energy expansion.

Finally, let us comment on the fact that we are using highly idealized
plates which provides perfect metallic boundary conditions. More
realistic plates would be associated with some cutoff $C<m$
representing the physical nature of the plates. More precisely, this
means that modes of the electromagnetic field with momenta $p\gtrsim
C$ are able to penetrate the plates and are not confined by them. This
situation was investigated in \cite{bordag-sch}. The effective theory
discussed in this paper can be viewed as a special case of this more
general situation with the external momenta $p$ restricted by $1/L\leq
p\ll C$. It is clear that effective field theory is powerful enough to
provide a description of this more realistic case too. That, however,
is beyond the scope of this paper.

{\em Note}: After this work was completed there appeared a paper by
K.\ Melnikov on the archives \cite{melnikov} which also deals with
radiative corrections to the Casimir energy and effective field
theory. Melnikov also finds an effective Lagrangian that reproduces
the BRW result. However, the term that is responsible for the
corrections in his Lagrangian is a bulk term proportional to $n_\mu
F^{\mu\rho}n^\nu F_{\nu\rho}$ (in our notation) with a coefficient
involving the plate separation $L$. To our minds, it would be
surprising if this term would be present in the correct effective
Lagrangian. There are two reasons for this: First, the operator
dimension of this term is $4$ which is the same as the leading Maxwell
term, and second, the parameter $L$ is a {\em global} property of the
system and is not {\em a priori} expected to occur in a local quantity
like the Lagrangian. To truly resolve the question of which effective
Lagrangian is correct it is necessary to calculate some other quantity
than the Casimir energy like for instance the propagator to order
$\alpha$ in both the full and effective theories. One should then be
able to determine which effective theory reproduces the result from
the full theory.

\noindent
\paragraph{Acknowledgments} One of us (F.R.) wants to thank C.P.\
Burgess for many useful discussions.

\end{document}